\documentclass[12pt]{article}
\usepackage{amsfonts}
\usepackage{amsmath}

\advance \topmargin by -\headheight
\advance \topmargin by -\headsep
     
\evensidemargin \oddsidemargin
\begin{document}

\topmargin -.6in
\def\rf#1{(\ref{eq:#1})}
\def\lab#1{\label{eq:#1}}
\def\nonu{\nonumber}
\def\br{\begin{eqnarray}}
\def\er{\end{eqnarray}}
\def\be{\begin{equation}}
\def\ee{\end{equation}}
\def\eq{\!\!\!\! &=& \!\!\!\! }
\def\ba{\be\begin{array}{c}}
\def\ea{\end{array}\ee}
\def\foot#1{\footnotemark\footnotetext{#1}}
\def\lb{\lbrack}
\def\rb{\rbrack}
\def\llangle{\left\langle}
\def\rrangle{\right\rangle}
\def\blangle{\Bigl\langle}
\def\brangle{\Bigr\rangle}
\def\llb{\left\lbrack}
\def\rrb{\right\rbrack}
\def\Blb{\Bigl\lbrack}
\def\Brb{\Bigr\rbrack}
\def\lcurl{\left\{}
\def\rcurl{\right\}}
\def\({\left(}
\def\){\right)}
\def\v{\vert}                     
\def\bv{\bigm\vert}               
\def\lskip{\vskip\baselineskip\vskip-\parskip\noindent}
\def\mskp{\par\vskip 0.3cm \par\noindent}
\def\sskp{\par\vskip 0.15cm \par\noindent}
\def\bc{\begin{center}}
\def\ec{\end{center}}

\def\tr{\mathop{\rm tr}}                  
\def\Tr{\mathop{\rm Tr}}                  
\makeatletter
\newcommand{\rd}{\@ifnextchar^{\DIfF}{\DIfF^{}}}
\def\DIfF^#1{%
   \mathop{\mathrm{\mathstrut d}}%
   \nolimits^{#1}\gobblespace}
\def\gobblespace{\futurelet\diffarg\opspace}
\def\opspace{%
   \let\DiffSpace\!%
   \ifx\diffarg(%
   \let\DiffSpace\relax
   \else
   \ifx\diffarg[%
   \let\DiffSpace\relax
   \else
   \ifx\diffarg\{%
   \let\DiffSpace\relax
   \fi\fi\fi\DiffSpace}
\newcommand{\deriv}[3][]{\frac{\rd^{#1}#2}{\rd #3^{#1}}}
\providecommand*{\dder}[3][]{%
\frac{\rd^{#1}#2}{\rd #3^{#1}}}
\providecommand*{\pder}[3][]{%
\frac{\partial^{#1}#2}{\partial #3^{#1}}}
\newcommand{\renewoperator}[3]{\renewcommand*{#1}{\mathop{#2}#3}}
\renewoperator{\Re}{\mathrm{Re}}{\nolimits}
\renewoperator{\Im}{\mathrm{Im}}{\nolimits}
\providecommand*{\iu}%
{\ensuremath{\mathrm{i}\,}}
\providecommand*{\eu}%
{\ensuremath{\mathrm{e}}}
\def\a{\alpha}
\def\b{\beta}
\def\c{\chi}
\def\d{\delta}
\def\D{\Delta}
\def\eps{\epsilon}
\def\vareps{\varepsilon}
\def\g{\gamma}
\def\G{\Gamma}
\def\grad{\nabla}
\newcommand{\h}{\frac{1}{2}}
\def\l{\lambda}
\def\om{\omega}
\def\s{\sigma}
\def\O{\Omega}
\def\p{\phi}
\def\vp{\varphi}
\def\P{\Phi}
\def\pa{\partial}
\def\pr{\prime}
\def\ti{\tilde}
\def\wti{\widetilde}
\def\lie{{\cal G}}

\newcommand{\cA}{\mathcal{A}}
\newcommand{\cB}{\mathcal{B}}
\newcommand{\cC}{\mathcal{C}}
\newcommand{\cD}{\mathcal{D}}
\newcommand{\cE}{\mathcal{E}}
\newcommand{\cF}{\mathcal{F}}
\newcommand{\cG}{\mathcal{G}}
\newcommand{\cH}{\mathcal{H}}
\newcommand{\cI}{\mathcal{I}}
\newcommand{\cJ}{\mathcal{J}}
\newcommand{\cK}{\mathcal{K}}
\newcommand{\cL}{\mathcal{L}}
\newcommand{\cM}{\mathcal{M}}
\newcommand{\cN}{\mathcal{N}}
\newcommand{\cO}{\mathcal{O}}
\newcommand{\cP}{\mathcal{P}}
\newcommand{\cQ}{\mathcal{Q}}
\newcommand{\cR}{\mathcal{R}}
\newcommand{\cS}{\mathcal{S}}
\newcommand{\cT}{\mathcal{T}}
\newcommand{\cU}{\mathcal{U}}
\newcommand{\cV}{\mathcal{V}}
\newcommand{\cW}{\mathcal{W}}
\newcommand{\cX}{\mathcal{X}}
\newcommand{\cY}{\mathcal{Y}}
\newcommand{\cZ}{\mathcal{Z}}
\newcommand{\nit}{\noindent}
\newcommand{\ct}[1]{\cite{#1}}
\newcommand{\bi}[1]{\bibitem{#1}}

\begin{center}
{\large\bf  Supersymmetry for integrable hierarchies }
\end{center}
\begin{center}
{\large\bf  on loop superalgebras  }
\end{center}
\normalsize
\vskip .4in

\begin{center}
 H. Aratyn

\par \vskip .1in \noindent
Department of Physics \\
University of Illinois at Chicago\\
845 W. Taylor St.\\
Chicago, Illinois 60607-7059\\
\par \vskip .3in

\end{center}

\begin{center}
J.F. Gomes, G.M. de Castro, M.B. Silka and A.H. Zimerman

\par \vskip .1in \noindent
Instituto de F\'{\i}sica Te\'{o}rica-UNESP\\
Rua Pamplona 145\\
01405-900 S\~{a}o Paulo, Brazil
\par \vskip .3in

\end{center}

\begin{abstract}
The algebraic approach is employed to 
formulate $N=2$ supersymmetry transformations in the context of
integrable systems based on loop superalgebras $\widehat{\rm sl}(p+1,p), \, p \ge 1$
with homogeneous gradation. 
We work with extended integrable hierarchies, which contain 
supersymmetric AKNS and Lund-Regge sectors.

We derive the one-soliton solution for $p=1$ which solves positive and 
negative evolution equations of the $N=2$ supersymmetric model.

\end{abstract}

\section{Introduction}

The well-known examples of the mKdV and sine-Gordon models \ct{kdv-sg} 
as well as
the AKNS and the Lund-Regge 
(Complex sine-Gordon) models \ct{akns-lr} can be understood as positive and negative flows 
belonging to the same integrable hierarchy conveniently classified in 
terms of underlying algebraic structure. 
The basic objects in an algebraic approach to integrable models are 
the loop algebra ${\widehat \lie}$ endowed with a grading operator $Q$
and a constant semisimple generator $E$. 
The grading decomposes the loop algebra into graded subspaces 
and specifies the space of physical fields.

There already exists an extensive literature devoted to construction of integrable models with 
extended supersymmetry based on superspace formalism, (see e.g.
\ct{Delduc-Gallot,Madsen:1999ta,Popowicz:1995jf,
Sorin:2002mq,Delduc:2002nd,Nissimov:2001cq,Ivanov:1999ab,Delduc:1999br,
Delduc:1996mx,Ivanov:1996av}).
Here, purely algebraic approach without superspace techniques is employed to construct a  class of integrable 
hierarchies based on loop superalgebras $\widehat{\rm sl}(p+1,p), \, p \ge 1$
with homogeneous gradation. 
Within this framework the symmetry structure of the 
underlying model is realized 
as a graded subalgebra of $\widehat{\rm sl}(p+1,p)$ defined as a centralizer $\cK$ 
of a semisimple element $E$ ($\cK= {\rm Ker} (E)$).
The positive and negative grade elements in the center of $\cK$ generate
positive and negative evolution flows of the extended hierarchy.
The positive flows are of the generalized  AKNS type, while the first
negative flow corresponds to the Lund-Regge (relativistic) model.
By comparing expressions for subsequent positive flows we derive the closed expression 
for the recursion operator $\cR$.

The elements in $\cK$ not residing in the center give rise to the non-abelian 
symmetry structure of the model.
In particular, $N=2$ supersymmetry transformations arise in this context
being generated by specific fermionic generators $f_{\pm} \in \cK$.
The supersymmetry algebra follows from the fact that the anti-commutator
of $f_{\pm}$ closes on $E$ \ct{Delduc-Gallot,Madsen:1999ta}

Furthermore, the algebraic approach allows a systematic construction of
Hamiltonians.
We derive the Lagrangian density for the second flow and the canonical (Dirac)
bracket which defines the first Poisson bracket structure of the model.
Higher brackets can be obtained by successive applications of recursion
operator.

We employ the dressing and vertex formalisms to construct 
soliton solutions. Explicit formulas for the one-soliton solution
are found  for $p=1$. The formula is verified to be a solution
of both the AKNS and Lund-Regge type of evolution equations.

This work completes the discussion started in \cite{Aratyn:2003ym},
where integrable models based on loop superalgebras
with principal gradation (like supersymmetric KdV model) were treated by 
algebraic methods.

Section 2, gives details of algebraic foundation of our construction.
We write equations of motion governed by the second flow 
for the $\widehat{\rm sl}(p+1,p)$ model for arbitrary $p$ generalizing 
the results of \ct{aratyn-rasinariu}.
Symmetries, conservation laws, recursion operator and the first 
Poisson bracket are presented within the algebraic framework in section 3.
In section 4,
we deal with the first negative flow of the extended hierarchy,
which is shown to be invariant under $N=2$ supersymmetry.  Next we write
explicit the Lund-Regge equations of motion (for the $p=1$) model 
using  the Gauss decomposition of  the zero grade subgroup element $B$.
In section 5, we explicit construct  one soliton solution by  using the
dressing and vertex operator techniques.

\section{General Construction of the Model}

Consider the graded Lie algebra ${\rm sl}(p+1, s+1)$  composed of  
bosonic (even) $\a_i, \b_t,\, i=1,\ldots, p, \;\; t=1, \ldots, s$ 
and fermionic (odd), $\g$, simple roots: 
\br
\a_i &=& e_i - e_{i+1}, \quad i=1,\ldots, p\nonu \\
\b_{t} &=& f_t - f_{t+1},  \quad t=1, \ldots, s \label{1}\\
\g &=& e_{p+1} - f_1 \nonu
\er
in terms of unit length vectors $e_r$, $f_a$  such that 
$e_r^2 = 1$ and $f_a^2 =-1$ with 
$r=1, \ldots,p+1$ and $a=1, \ldots,s+1$.
The remaining bosonic roots can be written  as
\br
\pm (\a_r + \cdots + \a_{q-1} )= \pm (e_r - e_q), \quad \pm (\b_a + \cdots + \b_{b-1} )= \pm (f_a - f_b), 
\label{2}
\er 
where $r,q = 1, \cdots, p+1$ and $a,b = 1, \cdots, s+1$.
The remaining fermionic roots have the form 
\br
\pm (\a_r + \cdots + \a_{p+1}+ \g + \b_1 +\cdots + \b_{a-1} )=\pm (e_r - f_a).   
\label{3}
\er
Consider the homogeneous gradation with the grading operator
$Q  = \l \dder{}{\l}$. The affine super Lie algebra 
$\widehat \lie = \widehat {{\rm sl}}(p+1, s+1)$ decomposes
according to $\widehat \lie = \oplus \lie_k$ with 
the graded subspaces $\lie_k$ that satisfy
$ [Q , \lie_k]  = k\lie_k$ with $k \in {\mathbb Z}$ and consist of elements
$X^{(k)}=\l^k X$ for $X \in {\rm sl} (p+1,s+1)$.

The key role in an algebraic approach to integrable models 
is played by the constant grade one generator 
\be
E= E^{(1)}
\label{4}
\ee
obtained by setting $n=1$ in the following expression for the semisimple
generators  of grade $n$
\[
E^{(n)} = \( \sum_{i=1}^{p} e_i - \sum_{a=1}^{s+1}f_a \) \cdot H^{(n)}  \, .
\]

The centralizer $\cK={\rm Ker} (E)$ is a subalgebra of
$\lie$ spanned by the Cartan subalgebra of ${\rm sl}(p+1, s+1)$ 
and the step operators associated to  roots 
\br
&&\pm (\a_i + \cdots + \a_j ) = \pm (e_i - e_j), \quad \pm (\b_a +\cdots + \b_{b-1}) = \pm (f_a - f_b),\nonu \\ 
&& \pm ( \a_i + \cdots \a_p +\g + \b_1 +\cdots + \b_{a-1})= \pm (e_i - f_a), 
\nonu \\
&&i,j =1, \cdots ,p, \quad a,b=1, \cdots ,s+1
\label{5}
\er
The image $\cI={\rm Im} (E)$ is then obtained as a linear combination
of step operators of 
bosonic roots
\[\pm (\a_i +\cdots \a_p )= \pm (e_{i} - e_{p+1}), \quad i=1,
 \cdots, p
\]
and fermionic roots  
\[ 
\pm (\g + \b_1 +\cdots + \b_{a-1})= \pm (e_{p+1} - f_a), \quad 
a = 1, \cdots, s+1\, ,
\]
respectively.

The fact that the element $E$ is semisimple insures that
$\lie= \cK \oplus \cI$. In addition, it also holds that
$\lb \cI , \cI \rb \subset \cK$.

Because supersymmetry requires equal number of bosons and fermions, 
we  impose from now on the condition $p=s+1$
and denote 
$\a_{p+1}=\g$, $\a_{p+2}=\b_1$, $ \a_{p+3}=\b_2, \cdots ,  \a_{2p}=\b_{p-1}$.

The super algebra $\widehat{{\rm sl}}(p+1, p)$ gives rise to a supersymmetric integrable model 
specified by element $E$ of eqn. (\ref{4}) and 
the following Lax operator :
\br
L = \pa_x + E + A_0 = \pa_x + {\cal A}_x\, ,
\label{7}
\er
where ${\cal A}_x=E + A_0 $ and 
\br
A_0 &=& \sum_{i=1}^p \(\bar  b_i E_{e_{i} - e_{p+1}}^{(0)}+   b_i E_{-(e_{i} - e_{p+1})}^{(0)}\) +
\sum_{a=1}^p \( \bar F_a E_{f_a - e_{p+1}}^{(0)}+   F_a E_{-(f_a - e_{p+1})}^{(0)} \), \label{8}  \\
&=& \sum_{i=1}^p \(\bar  b_i E_{\a_i +\cdots +\a_p}^{(0)}+   b_i E_{-(\a_i +\cdots +\a_p)}^{(0)}\) +
\sum_{a=1}^p \( \bar F_a E_{-(\a_{p+1} + \cdots + \a_{p+a})}^{(0)}+   
F_a E_{\a_{p+1} + \cdots + \a_{p+a}}^{(0)} \)\, . \nonu 
\er
lies in grade zero sector of $\cI$.

Now, we propose the zero curvature relation 
$ [ \pa_{t_n} + {\cal A}_{t_n}, \pa_x + \cA_x ]=0$.
We search for solution of the form 
\br
{\cal A}_{t_n} = D^{(n)} + D^{(n-1)}+ \cdots  D^{(0)} \, ,
\label{10} 
\er
where $ D^{(k)}\in \lie_k$. The  zero curvature relation  
decomposes grade by grade into   the following chain of equations,
\br
[E, D^{(n)}]&=&0,  \nonu \\
-\pa_x D^{(n)} + [A_0, D^{(n)}] + [E,  D^{(n-1)}]&=&0, \nonu \\
 \vdots&&\vdots  \label{11} \\
 -\pa_x D^{(1)}  + [A_0, D^{(1)}] + [E, D^{(0)}]&=&0, \nonu \\
-\pa_x D^{(0)}+ \pa_{t_2}A_0 + [A_0, D^{(0)}] &=&0 \, .
\nonu
\er
The top equation in (\ref{11}) implies that
$ D^{(n)} \in \cK$ and, consequently, we assume that
$D^{(n)}= E^{(n)} $.

Solving eqns. (\ref{11}) from the highest to the zero grade components we find for $t_n = t_2$, the  following
 equations of motion for the coefficients of $A_0$:
 \be \begin{split}
 \pa_{t_2}b_i &+ \pa^2_{x} b_i -2 \sum_{j=1}^{p} \( b_j\bar b_j +F_j\bar F_j  \) b_i =0,  \\
\pa_{t_2}\bar b_i &- \pa^2_{x} \bar b_i +2  \sum_{j=1}^{p} \( b_j\bar b_j +F_j\bar F_j \)\bar  b_i =0,  \\
\pa_{t_2}F_i &+ \pa^2_{x} F_i -2  \sum_{j=1}^{p} \( b_j\bar b_j +F_j\bar F_j \) F_i =0, \\
\pa_{t_2}\bar F_i &- \pa^2_{x} \bar F_i +2  \sum_{j=1}^{p} \( b_j\bar b_j +F_j\bar F_j \)\bar  F_i =0, 
\end{split}
\lab{a2}
\ee
where $i=1, \cdots , p $. These result generalizes equations of motion found
in \ct{AraRa} in the $p=1$ case.

\section{Symmetries and Conservation Laws}
The symmetries of integrable models will be defined below
in a systematic manner in the setting 
of a gauge  transformation  relating the  bare Lax ($\pa_x +E$) 
to the dressed Lax operator $L$  given by definition (\ref{7}) \cite{kluwer}.  

The dressing operator is defined by
\[
\Theta = e^{\theta^{(-1)} + \theta^{(-2)} + \cdots }, \qquad
\theta^{(-k)} \in \lie_{-k}
\]
and satisfies the  relation: 
\be
E = \Theta^{-1} \(  E + A_0 \)\Theta +   \Theta^{-1}\pa_x \Theta\, . 
\lab{14}
\ee
Decomposing \rf{14} according to the grading we find the following equations,
\be \begin{split}
A_0 + [E, \theta^{(-1)} ] &= 0,  \\
\pa_x \theta^{(-1)} + [E, \theta^{(-2)} ] +
\frac{1}{2}[A_0, \theta^{(-1)} ]     &= 0, \\
\pa_x \theta^{(-2)} + [E^{(1)}, \theta^{(-3)}] +\frac{1}{2}[A_0,
\theta^{(-2)}] + \frac{1}{12}[[A_0, \theta^{(-1)}],
\theta^{(-1)}] &= 0, \\
\pa_x \theta^{(-3)} + [E^{(1)}, \theta^{(-4)}] +\frac{1}{2}[A_0, , \theta^{(-3)}]+ \frac{1}{12}[[A_0, \theta^{(-1)}],
\theta^{(-2)}] \\+ \frac{1}{12}[[A_0, \theta^{(-2)}], \theta^{(-1)}]] & = 0 
\\
&  \vdots 
\end{split}
\label{15}
\ee
Decompose  $\theta^{(-i)}$ in its components in the centralizer $\cK$ and
image $\cI$ according to $\theta^{(-i)}= \theta^{(-i)}_{\cK} +
\theta^{(-i)}_{\cI}$.
Then the top eqn. (\ref{15}) yields 
\be
\theta^{(-1)}_{\cI} =\sum_{i=1}^p \(-\bar  b_i E_{\a_i +\cdots +\a_p}^{(0)}+   b_i E_{-(\a_i +\cdots +\a_p)}^{(0)}
- \bar F_a E_{-(\a_{p+1} + \cdots + \a_{p+i})}^{(0)}+   
F_a E_{\a_{p+1} + \cdots + \a_{p+i}}^{(0)} \), \lab{19a}
\ee
The second eqn. (\ref{15}) can be rewritten as 
\be \begin{split}
\pa_x \theta^{(-1)}_{\cK}+ \frac{1}{2} [ A_0, \theta^{(-1)}_{\cI}]
&= 0,  \\
\pa_x \theta^{(-1)}_{\cI}+[E, \theta^{(-2)}_{\cI}] + 
\frac{1}{2} [ A_0, \theta^{(-1)}_{\cK}]&= 0
\end{split}
\lab{19b}
\ee
and can be used to determine a local expression for $\theta^{(-2)}_{\cI}$.

An element $X_m \in {\rm Ker}(E)$ of $m$-th grade
generates a symmetry transformation via   \cite{kluwer}:
\be
 \d_{X_m} \Theta =  \(\Theta X_m \Theta^{-1}\)_{-}\, \Theta \, .
 \lab{transTheta}
\ee
We will derive the corresponding transformation of $A_0$ using 
that from the top relation in (\ref{15}) it follows that
\be
A_0 = \lb \theta^{(-1)}, E \rb = \( \lb \Theta , E \rb \)_0 \, ,
\lab{aztheta}
\ee
where the right hand
side contains projection on the zero grade.

Employing the definition \rf{transTheta} of the symmetry transformations
to the above relation we find:
\[ \begin{split}
 \d_{X_m}A_0  &= \( \lb \( \Theta X_m \Theta^{-1}\) _{-1} \Theta , E \rb
 \)_0 = \lb \( \Theta X_m \Theta^{-1}\) _{-1}, E \rb \\
 & = \( \lb \Theta X_m \Theta^{-1}, E \rb \)_0=
\( \big\lb \lb \Theta, E \rb \Theta^{-1}\,, \, \Theta X_m \Theta^{-1}
\big\rb \)_0 \, .
\end{split}
\]
{}From equation \rf{14} we find that 
\[ 
\lb \Theta, E \rb \Theta^{-1} = A_0+ \pa_x \Theta \, \Theta^{-1} 
\]
and, thus, 
transformation of $A_0$ can be given by a general gauge transformation
formula:
\be
\d_{X_n} A_0 = \( \big\lb A_0 + \pa_x \Theta \, \Theta^{-1} \, , \, \Theta X_n
\Theta^{-1} \big\rb \)_0 \, ,
\lab{dxna}
\ee
where $X_n$ is an element in $\cK$ of $n$-th grade.
For $m=0$, we find 
\be
\d_{X_0} A_0 = \left[ A_0 \, , \, X_0 \right].
\lab{18}
\ee
For $m=1$, eqn. \rf{dxna} yields,
 \be
 \d_{X_1}A_0 =  [A_0, [\theta^{(-1)}, X_1]]+[\pa_x \theta^{(-1)}, X_1] \,.
\lab{dxonea}
 \ee
In particular, for $X_1 =E$ we find that
$\d_{E}A_0  = \pa_x A_0 = \pa A_0/\pa t_1$, in accordance with the fact 
that the center of $\cK$ generates isospectral flows.

Define two odd elements of ${\rm Ker} (E)$ :
 \be
 f_{\pm} = \sum_{i=1}^p \(\eps^{(0)} E_{e_i-f_i}^{(0)} \pm \eps^{(1)}
 E_{-(e_i-f_i)}^{(1)}\)\, , 
 \lab{f}
\ee
containing grade $0$ and $1$ generators and Grassmannian parameters
$\eps^{(0)}$ and $\eps^{(1)}$. 
These two odd elements satisfy the graded commutation relations:
 \[
 \left[ f_{+},f_{-} \right] 
 = -2 \eps^{(0)}  \eps^{(1)} E, \quad \quad 
 \]
According to relation \rf{aztheta} the elements $f_{\pm} \in {\rm Ker} (E)$ 
give rise the symmetry transformations $\d_{f_{\pm}}$. 
These transformations satisfy the basic $N=2$ supersymmetry relations 
\[\lb \d_{f_{+}} , \d_{f_{-}} \rb A_0 =-2 \eps^{(0)}  \eps^{(1)} 
\pa_x A_0, \qquad 
\lb \d_{f_{\pm}}, \d_{f_{\pm}} \rb  A_0 =0\, .\]

For the particular case of $p=1$, i.e. ${\rm sl} (2,1)$, we find the  
 supersymmetry transformations  to be
\be \begin{split}
\d_{f_{\pm}} b_1 &= \pm \eps^{(1)} \( \pa_x F_1 -  b_1 \int \bar b_1 F_1  + F_1 \int (b_1 \bar b_1 + F_1 \bar F_1)\) ,  \\
\d_{f_{\pm}}\bar b_1 &= -\eps^{(0)} \bar F_1 \pm  \eps^{(1)}\bar b_1 \int \bar b_1 F_1 , \\
\d_{f_{\pm}}\bar F_1 &= \pm \eps^{(1)} \(\pa_x \bar b_1  - \bar b_1 \int (b_1 \bar b_1 + F_1 \bar F_1) + \bar F_1 \int \bar b_1 F_1 \), \\
\d_{f_{\pm}} F_1 &= \eps^{(0)} b_1 \mp \eps^{(1)} F_1 \int \bar b_1 F_1 , 
\end{split}
\label{a5}
\ee
These are indeed symmetries of the equations of motion \rf{a2}.  For the 
$\widehat{{\rm sl}} (p+1,p)$ case, the supersymmetry transformation  
generated by $f_{\pm}$ in \rf{f} are explicitly listed in the appendix A.

Let us now calculate the Hamiltonian densities. 
According to \cite{kluwer},
the Hamiltonian densities are given by 
\be
{\cal H}_n= -\tr \(E^{(0)} A^{(-n)} \) = \frac{1}{2} \sum_{k=0}^{n-1} \tr \(
A^{(-k)}A^{(1+k-n)}\)\, ,
\lab{17a}
\ee
where $A^{(-n)}$ are defined via
\[ 
\pa_x \Theta \, \Theta^{-1} = \sum_{k=1}^{\infty} A^{(-k)} \l^{-k} \,,
\]
where $\pa_x \Theta \, \Theta^{-1}$ is the quantity, which entered 
eqn. \rf{14}. 
The symbol $\tr$ in expression \rf{17a} denotes a trace for the 
super algebra ${\rm sl} (p+1,p)$.

Now, we explicitly work out the first few expressions. 
Inserting  $n=1$ in \rf{17a} we obtain:
\br
{\cal H}_1 &=& -\tr (E^{(0)} A^{(-1)} ) = \frac{1}{2}\tr ( A_0^2)
=\sum_{i=1}^{p}\( b_i \bar b_i + F_i \bar F_i \)
\label{18a}
\er
where $A_0 $ is given by (\ref{8}). In order to evaluate $  {\cal H}_2$
we use \rf{19a} to obtain 
\br
{\cal H}_2&=&  \tr \(A_0 \pa_x \theta^{(-1)}_{\cI}\)
= \sum_{i=1}^{p}\(\bar b_i \pa_x b_i + \bar F_i \pa_x F_i \)
\label{20a}
\er
For $n=3$, we find
\br
{\cal H}_3&=& -\frac{1}{2}\tr \( 2 A_0 A^{(-2)} + {A^{(-1)}}^2 \)
\er
where 
\br
A^{(-1)} = -\pa_x \theta^{(-1)}, \quad A^{(-2)} = \pa_x \theta^{(-2)} -\frac{1}{2} [\pa_x \theta^{(-1)}, \theta^{(-1)}]
\er
and, therefore,
\br
\frac{1}{3}{\cal H}_3 &=& \sum_{j=1}^{p} \pa_x \bar b_j \pa_x b_j + \sum_{a=1}^{p} \pa_x \bar F_a \pa_x F_a +
\sum_{i,j=1}^{p}\bar b_j b_j \bar b_i b_i + 
2  \sum_{i,b=1}^{p}\bar b_i b_i \bar F_b F_b \nonu \\
&+& 
\sum_{a,b=1}^{p}\bar F_a F_a \bar F_b F_b \,. 
\label{ham3} 
\er

\subsection{Poisson Structure of the second flow}
The equations of motion \rf{a2}  can be derived from the Lagrangian
density :
\br
{\cal L}_{t_2} &=&   \frac{1}{2} \sum_{i=1}^{p}  \pa_t  b_i \bar b_i -\frac{1}{2} \sum_{i=1}^{p} b_i \pa_t \bar b_i 
-  \frac{1}{2} \sum_{a=1}^{p}  \pa_t  F_a \bar F_a + \frac{1}{2} \sum_{a=1}^{p} F_a \pa_t \bar F_a  
-\sum_{j=1}^{p} \pa_x \bar b_j \pa_x b_j
\nonu \\
&-& \sum_{a=1}^{p} \pa_x \bar F_a \pa_x F_a -
\sum_{i,j=1}^{p}\bar b_j b_j \bar b_i b_i - 2  \sum_{i,b=1}^{p}\bar b_i b_i \bar F_b F_b - 
\sum_{a,b=1}^{p}\bar F_a F_a \bar F_b F_b
\label{ham}
\er
The canonical momenta are given by 
\br
P_{b_i} = {\frac{\d {\cal L}_{t_2}}{\d \dot b_i}} =  \frac{1}{2}\bar b_i, \quad 
P_{\bar b_i} = {\frac{\d {\cal L}_{t_2}}{\d \dot {\bar b}_i}} =-\frac{1}{2} b_i, \nonu \\
P_{F_a} = {\frac{\d {\cal L}_{t_2}}{\d \dot F_a}} =  -\frac{1}{2}\bar F_a, \quad 
P_{\bar F_a} = {\frac{\d {\cal L}_{t_2}}{\d \dot {\bar F}_a}} =-\frac{1}{2} F_a,
\label{p}
\er 
which satisfy the usual canonical Poisson brackets 
\br 
\{ P_{b_i}(x), b_j(y)\}_{P} = \{ P_{\bar b_i}(x), \bar b_j(y)\}_{P} = -\d_{ij} \d(x-y), \nonu \\
\{ P_{F_a}(x), F_b(y)\}_{P} = \{ P_{\bar F_a}(x), \bar F_b(y)\}_{P} =
-\d_{ab} \d(x-y) \, .
\nonu 
\er
The momenta dependence on the fields  in (\ref{p}) define the primary constraints
\br
 \Phi_{b_i} = P_{b_i} -\frac{1}{2}\bar b_i = 0, \quad \Phi_{\bar b_i} = P_{\bar b_i} +\frac{1}{2} b_i = 0, \nonu \\
\Phi_{F_a} = P_{F_a } +\frac{1}{2}\bar F_a = 0, \quad
\Phi_{\bar F_a} = P_{\bar F_a } +\frac{1}{2} F_a = 0, 
\label{pc}
\er
and accordingly the Poisson bracket algebra is given in terms of the Dirac brackets
\br
\{ {b_i}(x), \bar b_j(y)\}_{D} = -\d_{ij}\d(x-y), \quad \{ {F_a}(x), \bar
F_b(y)\}_{D} =-\d_{ab}\d(x-y)\, ,
\label{db}
\er
which together with the total Hamiltonian $H_3 = \int dx {\cal H}_3 $
obtained from (\ref{ham3}) reproduce
the equations of motion \rf{a2} via formula $\pa X / \pa t_2= \{X , H_3\}_D$.

A second  bracket characterized by $P_2$ can be obtained from the 
recursive relation 
\br 
P_2= {\cal R} P_1 \, ,
\er
where $P_1$ is the first bracket defined in (\ref{db}) 
and ${\cal R}$ denotes the recursion operator relating consecutive time evolutions, i.e.
\[
\pa_{t_m}A_0 = [E, \pa_x \pa_{t_{m-1}} A_0] -[A_0, \pa_x^{-1} [A_0,
[E,\pa_{{t_{m-1}}}A_0]]]= {\cal R} \pa_{t_{m-1}} A_0 \, .
\]
In components this recursion relation for flows reads as :
\be
\pa_{t_m} \xi_{i, l} = \sum_{k=1}^4 \sum_{j=1}^p 
\cR^{l,j}_{i,k} \pa_{t_{m-1}} \xi_{k,j} \,,
\quad i=1,{\ldots} ,4, \;\;\; l=1,{\ldots} ,p 
\lab{recurxi}
\ee
where the symbol $\xi_{i,j}$ represents all the fields contained in $A_0$
and introduced in equation \rf{19a} 
through $ \xi_{1, j} =\bar b_j$, $ \xi_{2,j}= b_j$, $\xi_{3,j}=F_j$
and $\xi_{4,j}=\bar F_j$.
For the case of ${\rm sl}(2,1)$ with $p=1$ one finds
the pseudo-differential operator expressions for the recursion operator
$\cR_{i,k} =\cR^{1,1}_{i,k}$:
\begin{alignat*}{2}
{\cal R}_{11} & =\pa_x -2 \bar b \pa_x^{-1} b + \bar F\pa_x ^{-1} F , &\quad  
\quad {\cal R}_{12} &= -2 \bar b \pa_x^{-1} \bar b , 
 \\
{\cal R}_{13} &=  \bar b \pa_x^{-1}  \bar F + \bar F \pa_x^{-1}\bar b , &\quad 
\quad {\cal R}_{14} &= -\bar b \pa_x ^{-1}  F, 
\\
{\cal R}_{21} &=2  b \pa_x ^{-1}  b,  &\quad  \quad
{\cal R}_{22} &= -\pa_x +2  b \pa_x^{-1} \bar b +F \pa_x^{-1} \bar F, 
\\
{\cal R}_{23} &=   b \pa_x ^{-1} \bar F, &\quad \quad 
{\cal R}_{24} &= F \pa_x^{-1} b + b \pa_x^{-1} F, 
\\
{\cal R}_{31} &= F\pa_x^{-1}   b  + b \pa_x ^{-1}  F ,&  \quad  \quad 
{\cal R}_{32} &=    F \pa_x ^{-1} \bar b , 
\\
{\cal R}_{33} &= -\pa_x + b \pa_x ^{-1} \bar b , &\quad \quad 
{\cal R}_{34}&= 0,   
\\
{\cal R}_{41}&= -\bar F \pa_x ^{-1} b_1 ,&  \quad \quad   
{\cal R}_{42}&= -\bar F \pa_x ^{-1} \bar b -  \bar b \pa_x ^{-1}\bar F, 
\\
{\cal R}_{43}&= 0, &  \quad \quad  
{\cal R}_{44}& = \pa_x -\bar b \pa_x ^{-1} b,  
\end{alignat*}
where $\pa_x $ and the pseudo-differential symbol $ \pa_x^{-1}$ are acting 
on all the fields appearing to their right.

In the appendix B we give the general expression for ${\cal R}$ 
in case of arbitrary ${\rm sl}(p+1,p)$. These expressions allow
an easy derivation of the second bracket.

\section{Relativistic Model}

Consider the negative grade time evolution equations given by
\br
\pa_{t_{-j}}A_0 - \pa_x (D^{(-1)}+D^{(-2)}+\cdots  +D^{(-j)}) - [ A_0 + E, D^{(-1)}+D^{(-2)}+\cdots  +D^{(-j)}]=0
\label{t-j}
\er
Here, we only consider $j=1$. We introduce notation
$z=t_{-1}$ and $\bar z=-x$. Thus, 
\[ {\pa}_z = \pder{}{t_{-1}} , \quad \bar \pa =\pder{}{\bar{z}} = - \pa_x\, ,\]
and the above evolution equation becomes
\[ 
\pa_z A_0 +\bar \pa D^{(-1)}- \lb A_0 +E, D^{(-1)} \rb=0\, ,
\]
with the following solution
\br
 D^{(-1)} = B E^{(-1)} B^{-1}, \quad A_0 = \bar \pa B \, B^{-1}
\label{lrakns}
\er
where $B$ is the group element of ${\rm sl}(p+1,p)$ satisfying
the Leznov-Saveliev's equations 
\be 
\bar{\partial}(B^{-1}\partial_z B)
+[E^{(-1)},B^{-1}E B]=0, \quad \quad 
\partial_z (\bar{\partial}B \, B^{-1}) 
+[BE^{(-1)} B^{-1},E ]=0 \, .
\lab{ls}
\ee
It is customary to think about coordinates $z, \bar z$ as 
light-cone coordinates. Such association allows us to interpret
the model as being relativistic.

The group element $B$ transforms under the symmetry transformations
\rf{transTheta} as
\be
\d_{X_n} B = -\(\Theta X_n \Theta^{-1}\)_{0}\, B\, , \quad n \ge 0\, .
\lab{dxnb}
\ee
For $n=0$ and $n=1$ the above relation takes form
\be
\d_{X_0}B = -X_0\, B, \quad \quad \d_{X_1}B = 
\big\lb X_1, \theta^{(-1)} \big\rb B \, .
\label{tr}
\ee
In particular, for $X_1=E$ the above definition yields
\[
\d_{E} B = -\(\Theta E \Theta^{-1}\)_{0}\, B= - \lb \theta^{(-1)}, E \rb B
= - A_0 B
\]
recalling that $A_0=-\pa_x B\, B^{-1}$ 
we obtain $\d_{E}B   = \pa_x B$, confirming again 
that $E$ generates the first positive isospectral flow $t_1=x$.

These transformations leave the above equations of motion \rf{ls} invariant.
In particular, the equations of motion \rf{ls} remain unchanged
under the $N=2$ supersymmetry transformations $\d_{f_{\pm}}$.

Notice that eqns. \rf{ls} yields chiral currents associated to the Kernel of $ E$, i.e. 
$ {\cal K} = \{ X \in {\rm sl}(p+1,p), [X, E]=0 \}$.
In order to make connection with the super AKNS model (\ref{8}), i.e. $A_0
\in \cI$ in (\ref{lrakns}),  
we impose the following subsidiary constraints 
\br
\tr \( X \bar{\partial}BB^{-1}\) = 
\tr \( X B^{-1} \partial_z B \) =0, \quad X \in {\cal K}
\label{constr}
\er

\subsection{Relativistic Super Lund-Regge Model}
In order to define the relativistic member of the so-called
 super Lund-Regge hierarchy 
 we consider the  zero grade group element $B\in G_0 = SL(2,1)$ 
 parametrized as
 \br
 B = e^{\tilde \chi E_{-\a_1}}e^{\tilde { f_1} E_{-\a_1-\a_2}}e^{\tilde { f_2} E_{\a_2}}e^{\frac{1}{2}\varphi_1 (\a_1+\a_2) \cdot H 
 + \frac{1}{2}\varphi_2 (\a_2) \cdot H}
 e^{\tilde {g_2} E_{-\a_2 }}e^{\tilde {g_1} E_{\a_1+\a_2}}e^{\tilde \psi E_{\a_1}}
 \lab{d1}
 \er
 where we are using the basis $E_{\pm \a_1}, E_{\pm \a_2},E_{\pm
 (\a_1+\a_2)}, \a_1 \cdot H $ and $\a_2 \cdot H$.  
 Changing to the natural variables
 \br
 \tilde \psi = \psi e^{-\frac{\varphi_{1}+\varphi_{2}}{2}}, \quad 
\tilde {g_{1}} =  g_{1}e^{-\frac{\varphi_{2}}{2}}, \quad
\tilde {f_{1}} =  f_{1}e^{-\frac{\varphi_{2}}{2}}\nonu \\
\tilde \chi =  \chi e^{-\frac{\varphi_{1}+\varphi_{2}}{2}}, \quad 
\tilde {g_{2}} =  g_{2}e^{-\frac{\varphi_{1}}{2}}, \quad
\tilde {f_{2}} =  f_{2}e^{-\frac{\varphi_{1}}{2}}
\lab{d2}
 \er
we find for the currents 
\br
J &=& B^{-1}\pa_z B = J_{-\a_1}E_{\a_1} + J_{-\a_2}E_{\a_2}+ J_{-\a_1-\a_2}E_{\a_1+\a_2}+ 
J_{(\a_1+ \a_2)\cdot H}\, (\a_1+\a_2)\cdot H \nonu \\
& -&  J_{ \a_2\cdot H} \,  \a_2 \cdot H
+ J_{\a_1}E_{-\a_1} + J_{\a_2}E_{-\a_2}+ J_{\a_1+\a_2}E_{-(\a_1+\a_2)},  \nonu \\
\bar J &=& \bar \partial B B^{-1} = \bar J_{-\a_1}E_{\a_1} + \bar J_{-\a_2}E_{\a_2}+ \bar J_{-\a_1-\a_2}E_{\a_1+\a_2}+ 
\bar J_{(\a_1+ \a_2)\cdot H}\,(\a_1+\a_2 )\cdot H \nonu \\
& -&  \bar J_{ \a_2\cdot H}\, \a_2\cdot H
+ \bar J_{\a_1}E_{-\a_1} + \bar J_{\a_2}E_{-\a_2}+ \bar J_{\a_1+\a_2}E_{-(\a_1+\a_2)}
\lab{d33}
 \er
with the coefficients $J_{-\a_1}, {\ldots} ,\bar J_{\a_1+\a_2}$ 
shown in Appendix D.

The constraints (\ref{constr})
\br
J_{\pm (\a_1+\a_2)} = \bar J_{\pm (\a_1+\a_2)} = J_{(\a_1 + \a_2)\cdot H} = \bar J_{(\a_1 + \a_2)\cdot H} = 
J_{ \a_2\cdot H} = \bar J_{ \a_2\cdot H} =0
\label{constr1}
\er 
lead to the following equations for the non local fields
\br
\pa_z {f}_{1}&=&\frac{1}{2}{f}_{1}\pa_z\varphi_{2}
+g_{2}[\pa_z\chi -\frac{1}{2}\chi (\pa_z\varphi_{1} +\pa_z\varphi_{2})], \nonu \\
\pa_z g_{1}&=&\psi\pa_z {f}_{2} +\frac{1}{2}g_{1} \pa_z\varphi_{2}
-\frac{1}{2}\psi {f}_{2}\pa_z\varphi_{1}, \nonu \\
 \bar{\partial} {f}_{1}&=&\chi\bar{\partial} g_{2}
+\frac{1}{2}{f}_{1} \bar{\partial}\varphi_{2} -\frac{1}{2}\chi
g_{2}\bar{\partial}\varphi_{1},\nonu \\
 \bar{\partial} g_{1}&=&\frac{1}{2}g_{1}\bar{\partial}\varphi_{2}
+{f}_{2}[\bar{\partial}\psi -\frac{1}{2}\psi
(\bar{\partial}\varphi_{1} +\bar{\partial}\varphi_{2})], \nonu \\
\pa_z\varphi_{1}&=&\frac{\psi [\pa_z\chi (1 +g_{2}{f}_{2})
+\frac{1}{2}\chi g_{2}\pa_z {f}_{2}]}{1 +\psi\chi (1
+\frac{5}{4}g_{2}{f}_{2})}, \nonu \\
\pa_z\varphi_{2}&=&\frac{\psi\pa_z\chi (1 +\frac{3}{2}g_{2}{f}_{2})
-g_{2}\pa_z {f}_{2} -\frac{1}{2}\psi\chi g_{2}\pa_z
{f}_{2}}{1 +\psi\chi (1 +\frac{5}{4}g_{2}{f}_{2})}, \nonu \\
 \bar{\partial}\varphi_{1}&=&\frac{\chi
[\bar{\partial}\psi (1 +g_{2}{f}_{2})
+\frac{1}{2}\psi\bar{\partial}g_{2}{f}_{2}]}{1 +\psi\chi (1
+\frac{5}{4}g_{2}{f}_{2})}, \nonu \\
 \bar{\partial}\varphi_{2}&=&\frac{\chi\bar{\partial}\psi
(1 +\frac{3}{2}g_{2}{f}_{2}) +(\frac{1}{2}\psi\chi
+1){f}_{2}\bar{\partial}g_{2}}{1 +\psi\chi (1
+\frac{5}{4}g_{2}{f}_{2})}
\lab{nonlocal}
\er
The equations of motion are given in the Leznov-Saveliev form \rf{ls},
after eliminating the non-local fields with help of equation \rf{nonlocal}.
In components they read
\br
(1 +g_{2}{f}_{2})[\bar{\partial}\pa_z\chi
-\frac{1}{2}\bar{\partial}\chi
(\pa_z\varphi_{1}+\pa_z\varphi_{2}) -\frac{1}{2}\chi
(\bar{\partial}\pa_z\varphi_{1}+\bar{\partial}\pa_z\varphi_{2})
+\frac{1}{2}(\bar{\partial}\varphi_{1}+\bar{\partial}\varphi_{2})\nonumber\\
(\pa_z\chi -\frac{1}{2}\chi (\pa_z\varphi_{1}
+\pa_z\varphi_{2}))]
+\bar{\partial}(g_{2}{f}_{2})[\pa_z\chi -\frac{1}{2}\chi
(\pa_z\varphi_{1} +\pa_z\varphi_{2})] +\chi=0, \label{eq1}
\er
\br
\bar{\partial}\pa_z g_{2}
+\frac{1}{2}\bar{\partial}g_{2}\pa_z\varphi_{1}
+\frac{1}{2}g_{2}\bar{\partial}\pa_z\varphi_{1}
-\frac{1}{2}\pa_z g_{2}\bar{\partial}\varphi_{1}
-\frac{1}{4}g_{2}\bar{\partial}\varphi_{1}\pa_z\varphi_{1} +(1
+\psi\chi)g_{2} =0
\er
\br
(1+g_{2}{f}_{2})[\pa_z\bar{\partial}\psi
-\frac{1}{2}\pa_z\psi
(\bar{\partial}\varphi_{1}+\bar{\partial}\varphi_{2})
-\frac{1}{2}\psi
(\pa_z\bar{\partial}\varphi_{1}+\pa_z\bar{\partial}\varphi_{2})
+\frac{1}{2}(\pa_z\varphi_{1}+\pa_z\varphi_{2})\nonumber\\(\bar{\partial}\psi
-\frac{1}{2}\psi (\bar{\partial}\varphi_{1}
+\bar{\partial}\varphi_{2}))]
+\pa_z(g_{2}{f}_{2})[\bar{\partial}\psi -\frac{1}{2}\psi
(\bar{\partial}\varphi_{1} +\bar{\partial}\varphi_{2})] +\psi=0
\er
\br
\pa_z\bar{\partial} {f}_{2}
+\frac{1}{2}\pa_z {f}_{2}\bar{\partial}\varphi_{1}
+\frac{1}{2}{f}_{2}\pa_z\bar{\partial}\varphi_{1}
-\frac{1}{2}\bar{\partial}{f}_{2}\pa_z\varphi_{1}
-\frac{1}{4}{f}_{2}\pa_z\varphi_{1}\bar{\partial}\varphi_{1}
+(1 +\psi\chi){f}_{2} =0
\label{eq4}
\er
where the non local fields $\varphi_1$ and $\varphi_2$ are given 
in equation \rf{nonlocal}.

\subsection{Connection between AKNS and Lund-Regge Variables}
On basis of (\ref{lrakns}) 
we find the following non-local relations between the AKNS and 
the relativistic Lund-Regge variables :
\br
\bar b_1 = \bar J_{-\a_1} &=& \frac{e^{\frac{1}{2}(\varphi_1 +\varphi_2)}}
{1+f_2g_2}
  \( \bar \pa \psi - \frac{1}{2}\psi (\bar \pa \varphi_1 + \bar \pa \varphi_2 )\), \nonu \\
F_1 = \bar J_{-\a_2} &=&e^{-\frac{1}{2}\varphi_1}\( \bar \pa f_2 + \frac{1}{2} f_2 \bar \pa \varphi_1 \), \nonu \\
b_1 = \bar J_{\a_1} & = & e^{-\frac{1}{2}(\varphi_1 +\varphi_2)}\( \bar \pa \chi +
\frac{1}{2}\chi (\bar \pa \varphi_1 +\bar \pa \varphi_2)-\chi f_2 \bar \pa g_2 -
\frac{1}{2}\chi  \bar \pa \varphi_1 g_2 f_2 + e^{\frac{1}{2}\varphi_1}f_1 \bar J_{-\a_2} \right. \nonu \\
&-& \left. \chi^2 e^{-\frac{1}{2}(\varphi_1 +\varphi_2)}\bar J_{-\a_1}\),\nonu \\
\bar F_1 = \bar J_{\a_2} &=& e^{\frac{1}{2}\varphi_1}\( \bar \pa g_2 -\frac{1}{2}g_2 \bar \pa \varphi_1 +
 e^{-\frac{1}{2}(\varphi_1 +\varphi_2)}f_1 \bar J_{-\a_1} \)
\label{rem}
\er
Notice that they are all non local fields in terms of the Lund-Regge 
variables due to the presence of $\varphi_1, \varphi_2$
and $f_1$ defined by \rf{nonlocal}. 
Also, we have seen that in terms of the AKNS variables, the models are 
invariant under  supersymmetry transformations (\ref{a5}). 

\section{Soliton Solutions}

\subsection{Dressing Transformations and Vertex Operators}

The dressing transformation and the vertex operators method represents a
powerful tool for the construction of solitons solutions of integrable models.
The dressing transformation  relates two solutions of the equations of motion written in the zero curvature 
representation.  In particular, it
relates  the vacuum and the 1-soliton solutions by a gauge transformation,
\br
 {\cal{A}}_{\mu}= \Theta_{\pm} {\cal{A}}_{\mu}^{ vac}\Theta_{\pm}^{-1} - \( \pa_{\mu}\Theta_{\pm}\) \Theta_{\pm}^{-1}
 \label{dress}
 \er
where 
\br
\Theta_- = e^{t(-1)+\cdots} , \quad \quad \Theta_+ =  e^{v(0)}e^{v(1)}\cdots
\label{asoliton} 
\er
The zero curvature representation implies for pure gauge solutions: 
\br
{\cal{A}}_{\mu}^{ vac }= -\pa_{\mu}T_0 T^{-1}_0 , \quad \quad {\cal{A}}_{\mu}= -\pa_{\mu}T T^{-1}
\label{t0}
\er
which allows the following relation 
\br
T = \Theta_{\pm} T_0,\; \; i.e. \; \; \Theta_+ T = \Theta_- T_0 g\, ,
\label{4.3}
\er
where $g \in \hat {G}$ is an arbitrary constant element of the corresponding affine group.  Suppose 
$T_0 $ represents  the vacuum solution, 
\br
T_0 = \exp (-t_n E^{(n)}) \exp (- x E) , \qquad n>1
\label{4.4}
\er
i.e.,
\br 
{\cal{A}}_{t_n}^{  vac }= E^{(n)}, \quad \quad  {\cal{A}}_{x}^{  vac }= E \, .
\label{avac}
\er
As consequence of (\ref{dress}) with (\ref{avac}) and (\ref{asoliton}) we 
can determine $\Theta_{\pm}$. 
Consider for instance eqn. (\ref{dress}) for $A_x$ and $\Theta_-$.   
It determines its zero grade component
$t(-1)$  through  
\br
A_0 = [t(-1), E]
\label{e}
\er
Notice, that if we enlarge the loop algebra by adding central terms,
consistency requires the introduction of a additional field $\nu$
associated with the extension $A_0 \rightarrow A_0 + \pa_x \nu \hat c$ in constructing 
the Lax operator.  For the specific ${\rm sl}(2,1)$  model:
\br
A_0 = \bar{b}_1 E_{\a_1}^{(0)} + b_1 E_{-\a_1}^{(0)} + F_1 E_{\a_2 }^{(0)}
+ \bar F_1 E_{- \a_2 }^{(0)} + \pa_x \nu \hat c
\label{laxsl21}
\er 
we find 
\br
t(-1) = -\bar{b}_1 E_{\a_1}^{(-1)} + b_1 E_{-\a_1}^{(-1)} +
 F_1 E_{\a_2 }^{(-1)}- \bar F_1 E_{- \a_2 }^{(-1)} + \pa_x \nu \frac{1}{2}(\a_1+\a_2)\cdot H^{(-1)}
\label{t-1sl21}
\er
From (\ref{4.3}) we find the solution
\br
\tau_0 &\equiv & e^{\nu } = < \l_0|  T_0 g T_0^{-1} |\l_0 >, \nonu \\
\tau_1 &\equiv & \bar b_2  \tau_0 = < \l_0| E_{-\a_1}^{(1)} T_0 g T_0^{-1} |\l_0 >, \nonu \\
\tau_2 &\equiv & b_2  \tau_0 = -< \l_0| E_{\a_1}^{(1)} T_0 g T_0^{-1} |\l_0 >, \nonu \\
\tau_3 &\equiv &  F_1  \tau_0 = < \l_0| E_{-\a_2}^{(1)} T_0 g T_0^{-1} |\l_0 >, \nonu \\
\tau_1 &\equiv &  \bar  F_1 \tau_0 = < \l_0| E_{\a_1}^{(1)} T_0 g T_0^{-1} |\l_0 >, \nonu \\
\label{tau}
\er

For the relativistic model let us consider eqn. (\ref{dress}) for $A_x$ and $\theta_+$.   We find
\br
A_0  = -\pa_x e^{v(0)}\, e^{-v(0)}
\er
Since,for the affine case, $A_0 = - \pa_x BB^{-1} - \pa_x \nu \hat c$,
\br
e^{v(0)} = B e^{\nu \hat c}
\er 
From (\ref{4.3}) we find the solution
\br
<\l_r |Be^{\nu \hat c} |\l_t> = <\l_r | T_0 g T_0^{-1}|\l_t>
\er
We now chose $   |\l_r>, \;   |\l_t> $ to be the following weight states
\br 
|\rho >=  |\frac{1}{2}\a_1 >, \quad \quad |\eps>=  |-\frac{1}{2}\a_1- \a_2 >
\label{states}
\er
and obtain
\br
<\rho |Be^{\nu \hat c} |\rho> &=& \tau_0 e^{\frac{1}{2}(\varphi_1 +\varphi_2)} =<\rho |T_0 g T_0^{-1} |\rho>, \nonu \\ 
<\eps |Be^{\nu \hat c} |\eps> &=& \tau_0 e^{\frac{1}{2}(\varphi_1 -\varphi_2)} =<\eps |T_0 g T_0^{-1} |\eps>, \nonu \\ 
<\rho |Be^{\nu \hat c} E_{-\a_1-\a_2}|\rho> &=&\frac{1}{2} \tilde g_1\tau_0 e^{\frac{1}{2}(\varphi_1 +\varphi_2)} 
=<\rho |T_0 g T_0^{-1} E_{-\a_1-\a_2}|\rho>, \nonu \\ 
<\rho |Be^{\nu \hat c} E_{\a_2}|\rho> &=&-\frac{1}{2} \tilde g_2\tau_0 e^{\frac{1}{2}(\varphi_1 +\varphi_2)} 
=<\rho |T_0 g T_0^{-1} E_{\a_2}|\rho>, \nonu \\
<\rho |Be^{\nu \hat c} E_{-\a_1}|\rho> &=&\tau_0 e^{\frac{1}{2}(\varphi_1 +\varphi_2)} \( \tilde \psi +\frac{1}{2}\tilde g_2 \tilde g_1\) 
=<\rho |T_0 g T_0^{-1} E_{-\a_1}|\rho>, \nonu \\
\rho |E_{\a_1+\a_2}Be^{\nu \hat c} |\rho> &=&\frac{1}{2} \tilde f_1\tau_0 e^{\frac{1}{2}(\varphi_1 +\varphi_2)} 
=<\rho |E_{\a_1+\a_2}T_0 g T_0^{-1} |\rho>, \nonu \\ 
<\rho |E_{-\a_2}Be^{\nu \hat c} |\rho> &=&-\frac{1}{2} \tilde f_2\tau_0 e^{\frac{1}{2}(\varphi_1 +\varphi_2)} 
=<\rho |E_{-\a_2}T_0 g T_0^{-1} |\rho>, \nonu \\
<\rho |E_{\a_1}Be^{\nu \hat c} |\rho> &=&\tau_0 e^{\frac{1}{2}(\varphi_1 +\varphi_2)} \( \tilde \psi +\frac{1}{2}\tilde g_2 \tilde g_1\) 
=<\rho |E_{\a_1}T_0 g T_0^{-1} |\rho>, \nonu \\
\label{sol}
\er

\subsection{Vertex Operators and Solutions }

In order to evaluate the matrix elements in the r.h.s. of eqns. (\ref{tau})
and  (\ref{sol}) it is instructive 
to introduce the eigenvectors  ($\cF (\g )$) of $E^{m}$, i.e.,
\begin{equation}
\lbrack E^{(m)},{\cal F}_i(\gamma )]=f^{m}_i(\gamma ){\cal F}_i(\gamma ), \; \;\;
\label{fgamma}
\end{equation}
such that we can classify the soliton solutions in terms of the constant group element  
$g = e^{\cF_1(\gamma_1 )}e^{\cF_2(\gamma_2 )} \cdots $
and hence 
\br 
T_0 g T_0^{-1}= \exp \( \rho_1 (\gamma_1) {\cal F}_1(\gamma_1 )+  \rho_2 (\gamma_2) {\cal F}_2(\gamma_2 ) + \cdots \)
\label{t0g}
\er
where $\rho_i (\gamma_i) = \exp \( -t_n f^{n}_i -  x f^{1}_i \)$.

For $\lie ={\rm sl}(2,1)$ model, $E^{(1)} = (\a_1+\a_2)\cdot H^{(1)}$ and its eigenstates are given by
\be \begin{split}
{\cal F}_{1}(\g ) &= \sum_{n\in {\mathbb Z}} \( E_{ \a_1}^{(n)} + a_{1}E_{- \a_2}^{(n)}\) \g^{-n}, \\
{\cal F}_{2}(\g ) &= \sum_{n\in {\mathbb Z}} \( E_{ -\a_1}^{(n)} + a_{2}E_{ \a_2}^{(n)}\) \g^{-n}, 
\end{split}
\lab{4.12}   
\ee
with eigenvalues $f_{1}^m(\g ) =  \g^m, \quad f_{2}^m(\g ) = -\g^{-m}$.  
We now consider soliton solutions associated to $g = \exp ( {\cal F}_{1}(\g_1 ) ) \exp ({\cal  F}_{2}(\g_2 ) )$ and 
\br
T_0 g T_0^{-1}=\( 1+ \rho_1 (\gamma_1){\cal  F}_{1}(\g_1 ) \)\( 1+ \rho_2 (\gamma_2){\cal  F}_{2}(\g_2 ) \)
\label{ver}   
\er
leading to the following matrix elements
\br
\tau_0 &=& <0| T_0 g T_0^{-1}|0>= 1+ \Gamma \rho_1^{\pr}\rho_2^{\pr}, \nonu \\
\tau_1 &=& <0|E_{-\a_1}^{(1)} T_0 g T_0^{-1}|0>= -\g_2 \rho_2^{\pr}, \nonu \\
\tau_2 &=& <0| T_0 g T_0^{-1}E_{\a_1}^{(-1)}|0>= \g_1 \rho_1^{\pr}, \nonu \\
\tau_3 &=& <0|E_{-\a_2}^{(1)} T_0 g T_0^{-1}|0>= -a_2\g_2 \rho_2^{\pr}, \nonu \\
\tau_4 &=& <0| T_0 g T_0^{-1}E_{\a_2}^{(-1)}|0>= a_1\g_1 \rho_1^{\pr}
\label{solmatr}
\er
where $b = a_1 a_2$,  $ \Gamma_0 =  \frac{{\g_1 \g_2}}{(\g_1 - \g_2)^2} $ and $\Gamma = (1-b)\Gamma_0$. 
We therefore find the solution for the non relativistic model (\ref{13a}),
\br
\bar b_1 &=& \frac{\g_1 \rho_1^{\pr}}{1 + \Gamma \rho_1^{\pr}\rho_2^{\pr}}, \quad 
 b_1 = -\frac{\g_2 \rho_2^{\pr}}{1 + \Gamma \rho_1^{\pr}\rho_2^{\pr}}, \nonu \\
F_1 &=& -a_2\frac{\g_2 \rho_2^{\pr}}{1 + \Gamma \rho_1^{\pr}\rho_2^{\pr}}, \quad
\bar F_1 = a_1\frac{\g_1 \rho_1^{\pr}}{1 + \Gamma \rho_1^{\pr}\rho_2^{\pr}}
\label{solnrel}
\er
where for the non relativistic model 
 $\rho_1^{\pr} (\gamma_1)=e^{-{{t_2} {\g_1}^2} -\g_1 x}, \quad \rho_2^{\pr} (\gamma_1)=e^{{{t_2} {\g_2}^2} +\g_2 x}$.

For the relativistic model described by equations of motion (\ref{eq1})-(\ref{eq4}) we find the following matrix elements
\be \begin{split}
<\rho | T_0 g T_0^{-1}|\rho >&= 1+ \rho_1 \rho_2 {\frac{\g_1}{(\g_1 -\g_2)^2}}\( \g_1 - \frac{1}{2} (\g_1 +\g_2)b\),  \\
<\eps | T_0 g T_0^{-1}|\eps >&= 1+ \rho_1 \rho_2 {\frac{\g_1}{(\g_1 -\g_2)^2}}\( \g_1 + \frac{1}{2} (\g_1 -3\g_2)b\), \\
<\rho | T_0 g T_0^{-1}E_{-\a_1}^{(0)}|\rho >&=  \rho_1, \\ 
<\rho |E_{\a_1}^{(0)} T_0 g T_0^{-1}|\rho >&=  \rho_2,  \\ 
<\rho | T_0 g T_0^{-1}E_{\a_2}^{(0)}|\rho >&=  -\frac{1}{2}a_1\rho_1, \\
<\rho |E_{-\a_2}^{(0)} T_0 g T_0^{-1}|\rho >&=  -\frac{1}{2}a_2\rho_2, \\
<\rho | T_0 g T_0^{-1}E_{-\a_1-\a_2}^{(0)}|\rho >&=
\frac{1}{2}a_1\rho_1\rho_2  {\frac{\g_1}{(\g_1-\g_2)}},  \\
<\rho |E_{\a_1+\a_2}^{(0)} T_0 g T_0^{-1}|\rho >&=  \frac{1}{2}a_1\rho_2
{\frac{\g_1}{(\g_1-\g_2)}}, 
\end{split}
\label{solrelmat}
\ee
which allow us to determine the solution
\be \begin{split}
 e^{\frac{1}{2}(\varphi_1 + \varphi_2)} &= {\frac{1+\frac{\g_1}{ \g_2}\Gamma_0
 \(1-b {\frac{(\g_1+\g_2)}{2\g_1}}\)\rho_1 \rho_2} {\tau_0}}, \\
 e^{\frac{1}{2}(\varphi_1 - \varphi_2)} &={\frac{1+\Gamma_0 \(1+b {\frac{(\g_1-3\g_2)}{2\g_2}}\)\rho_1 \rho_2}{\tau_0}} 
= 1- {\frac{b}{ 2}}{\frac{(\g_2 -\g_1)\Gamma_0 \rho_1 \rho_2}{\g_2 (1+\Gamma_0 \rho_1 \rho_2)}}, \\
\psi &= {\frac{\rho_1}{\tau_0}}\( 1- {\frac{b\g_1 \rho_1
\rho_2}{2(\g_1-\g_2)(1+{\frac{\g_1}{\g_2}}\Gamma_0 \rho_1 \rho_2)}}\), \\
 \chi &= {\frac{\rho_2}{\tau_0}}\( 1- {\frac{b\g_1 \rho_1
 \rho_2}{2(\g_1-\g_2)(1+{\frac{\g_1}{\g_2}}\Gamma_0 \rho_1 \rho_2)}}\), \\
 g_1 &= a_2{\frac{\g_1 \rho_1 \rho_2}{(\g_1-\g_2)\tau_0}}e^{-\frac{1}{2}\varphi_1}, \quad \quad 
  f_1 = a_1{\frac{\g_1 \rho_1 \rho_2}{(\g_1-\g_2)\tau_0}}e^{-\frac{1}{2}\varphi_1}, \\
 g_2 &= a_1{\frac{ \rho_1 }{\tau_0}}e^{-\frac{1}{2}\varphi_2}, \quad \quad
  f_2 = a_2{\frac{ \rho_2}{\tau_0}}e^{-\frac{1}{2}\varphi_2}, 
\label{solrel}
\end{split}
\ee
From the solutions given in (\ref{solrel}) it is easy to verify the consistency of
 the constraints (\ref{constr}) (or \rf{nonlocal}). Let
us take for instance 
\br
\bar \pa f_1 = \chi \bar \pa g_2 +\frac{1}{2}f_1 {\bar \pa \varphi_2} - \frac{1}{2}g_2 \chi \bar \pa \varphi_1
\label{check}
\er
Substituting (\ref{solrel}) in both sides of eqn. (\ref{check}) and 
identifying $t_{-1}=z,\;\; x = -\bar z$ 
using the second eqn. (\ref{solrel}) together with the fact that $a_1 b = a_2 b =0$ we 
find that the constraint  (\ref{check}) is satisfied.  The same can be shown for all 
eqns.  \rf{nonlocal} ( i.e. all constraints (\ref{constr}) are satisfied).

Finally we verify the connection between the non-relativistic and the relativistic variables given by eqns. (\ref{rem}).  Substituting the
solutions (\ref{solrel}) in  (\ref{rem}) and comparing with (\ref{solnrel}) we conclude that they agree if we associate 

\br
\g^2 t_2 \rightarrow {\frac{z}{\g}} ={\frac{t_{-1}}{\g}} , \quad \quad  x
\rightarrow - \bar {z}
\er
which implies $\rho(\g) \rightarrow \rho^{\pr}(\g)$.

\section*{\sf Acknowledgments}
H.A., G.M.C. and M.B.S. acknowledges support from Fapesp.
H.A. thanks the IFT-UNESP Institute
for its hospitality. JFG and AHZ thank CNPq for a partial support.

\appendix 
\section{Supersymmetry transformation for $\widehat{{\rm sl}}(p+1,p)$}

Let $f_{\pm}$ be the supersymmetry generators from equation \rf{f}.
Using the symmetry transformations given in equations
\rf{18} and \rf{dxonea} and 
\br
\theta^{(-1)}_{I} &=&  \sum_{i=1}^p \( -\bar b_i E^{(-1)}_{(e_i -e_{p+1})}  + b_i E^{(-1)}_{-(e_i -e_{p+1})} \) \nonu \\
&+&\sum_{a=1}^p \( -\bar F_a E^{(-1)}_{(f_a -e_{p+1})}  + F_a E^{(-1)}_{-(f_a -e_{p+1})} \)
\label{b3}
\er 
and
\br
\pa_x \theta ^{(-1)}_{K} &=& -\sum_{j=1, \; j \neq i}^p \sum_{i=1}^p \bar b_j b_i E^{(-1)}_{(e_i -e_{j})} - 
\sum_{j=1}^p \bar b_j b_j (e_j-e_{p+1})\cdot H^{(-1)} \nonu \\
&-& \sum_{j=1}^p \sum_{a=1}^p \bar b_j F_a E^{(-1)}_{(e_i -f_a)} 
- \sum_{j=1}^p \sum_{a=1}^p  b_j \bar F_a E^{(-1)}_{-(e_i -f_a)}\nonu \\
&-& \sum_{a=1, \; a \neq b}^p \sum_{b=1}^p \bar F_a F_b E^{(-1)}_{(f_a -f_b)}
+ \sum_{a=1}^p \bar F_a F_a{(f_a -e_{p+1})}\cdot H^{(-1)}
\label{b4}
\er
obtained from equation (\ref{15}) we find
\br
\d_{f_\pm}b_r &=&  \pm \( \pa_x F_r - \sum_{j=1}^p \(b_j \int \bar b_j F_r  - F_j \int  (b_r \bar b_j 
+  F_r \bar F_j ) \)\)  \eps^{(1)} , \nonu \\
\d_{f_{\pm}} \bar b_r &=&  - \bar F_r\eps^{(0)} \pm \( \sum_{j=1}^{p} 
\bar b_j \int \bar b_r F_j   \)\eps^{(1)} , \nonu \\
\d_{f_\pm}\bar F_r &=&
\pm \( \pa_x \bar b_r - \( \sum_{j=1}^p  \bar b_j \int  (b_j \bar b_r 
+   F_j  \bar F_r)\) + \bar F_r \int  \bar b_r  F_r \) \eps^{(1)}  
 , \nonu \\
\d_{f_\pm} F_r &=&  b_r \eps^{(0)}
\mp \sum_{j=1}^p F_j \int \bar b_j F_r  \eps^{(1)}  
\label{b5}
\er 
\section{Recursion Operator for ${\widehat {\rm sl}}(p+1,p)$}
Using the same notation as in equation \rf{recurxi} we write down 
expression for the recursion pseudo-differential operator $\cR^{lj}_{i,k}$
valid for a general case of  ${\widehat {\rm sl}}(p+1,p)$.
Here, the $l,j=1,{\ldots} ,p$ and indices $i,k=1,2,3,4$ label 
the four modes $\bar b_j, b_j, F_j, \bar F_j$.
\[ \begin{split}
 {\cal R}_{11}^{lj} &= \d_{lj}\(\pa_x -\sum_{k} \bar b_k \pa_x ^{-1} b_k + 
 \sum_{b}\bar F_b\pa_x ^{-1} F_b \)  -\bar b_l \pa_x ^{-1} b_j ,  \\ 
{\cal R}_{12}^{lj}   &= - \bar b_j \pa_x ^{-1} \bar b_l -  \bar b_l \pa_x ^{-1} 
\bar b_j ,  \\
{\cal R}_{13}^{lj}   &= \bar b_l \pa_x ^{-1}  \bar F_j + 
\bar F_j \pa_x ^{-1}\bar b_l ,  \\ 
{\cal R}_{14}^{lj}  &= -\bar b_l \pa_x ^{-1}  F_j ,  \\
{\cal R}_{21}^{lj}  &=     b_j \pa_x ^{-1}  b_l +  b_l \pa_x ^{-1}  b_j  ,   \\
 {\cal R}_{22}^{lj}  &=   \d_{lj}\( -\pa_x + \sum_{k}  b_k \pa_x ^{-1} \bar b_k +\sum_{b} F_b \pa_x ^{-1} \bar F_b \) 
 +   b_l \pa_x ^{-1} \bar b_j  ,  \\
 {\cal R}_{23}^{lj}   &=   b_l \pa_x ^{-1} \bar F_j ,  \\ 
 {\cal R}_{24}^{lj}   &=  F_j \pa_x ^{-1} b_l  + b_l \pa_x ^{-1}F_j ,  \\
 {\cal R}_{31}^{lj}    &=   F_l\pa_x ^{-1}   b_j + b_j \pa_x ^{-1}    F_l ,   \\
 {\cal R}_{32}^{lj}    &=  F_l \pa_x ^{-1} \bar b_j  ,  \\
 {\cal R}_{33}^{lj}  &=  \d_{lj}\( -\pa_x + \sum_{k} b_k \pa_x ^{-1} \bar b_k  
 +\sum_{b} F_b \pa_x ^{-1} \bar F_b \) 
 -    F_l \pa_x ^{-1} \bar F_j                 \\
 {\cal R}_{34}^{lj}  &=  F_l \pa_x ^{-1} F_j -F_j \pa_x^{-1} F_l,   \\
 {\cal R}_{41}^{lj}  &= -\pa_x ^{-1} b_j \bar F_l,   \\  
 {\cal R}_{42}^{lj}   &= -\bar F_l\pa_x ^{-1} \bar b_j  -  
 \bar b_j \pa_x ^{-1}\bar F_l \cdot  ),   \\
 {\cal R}_{43}^{lj}  &=  -\bar F_j \pa_x ^{-1} \bar F_l  + \bar F_l \pa_x ^{-1} \bar F_j    \\ 
 {\cal R}_{44}^{lj}   &=  \d_{lj}\( \pa_x - \sum_{k} \bar b_k \pa_x ^{-1} b_k   +\sum_b  \bar F_b \pa_x ^{-1}  F_b \) 
 - \bar F_l \pa_x ^{-1}  F_j .
\end{split}
\]

\section{Case of $\widehat{{\rm sl}} (3,2)$}

Consider the $\widehat{{\rm sl}} (3,2)$ case where 
\br
A_0 &=&  b_1 E_{-\a_1-\a_2}^{(0)}+  \bar b_1 E_{\a_1+\a_2}^{(0)}+ 
b_2 E_{-\a_2}^{(0)}+  \bar b_2 E_{\a_2}^{(0)} \nonu \\
& +& F_1 E_{\a_3}^{(0)}+  \bar F_1 E_{-\a_3}^{(0)} +
 F_2 E_{\a_3+\a_4}^{(0)}+  \bar F_2 E_{-\a_3-\a_4}^{(0)}
\label{9} 
\er
and propose the zero curvature representation 
\br
\pa_{t_2} A_0 - \pa_x \( D^{(0)} + D^{(1)} + D^{(2)} \)  - [ A_0 + E,  D^{(0)} + D^{(1)} + D^{(2)}] =0
\lab{10a} 
\er
from where we find the following solutions for equations (\ref{11}),
\br
D^{(2)} &=& (D^{(2)})_{Ker} = (e_1+e_2-f_1-f_2)\cdot H^{(2)}, \nonu \\
D^{(1)} &=& (D^{(1)})_{IM} = b_1 E_{-\a_1-\a_2}^{(1)}+   b_2 E_{-\a_2}^{(1)}+ 
\bar b_1 E_{\a_1+\a_2}^{(1)}+  \bar b_2 E_{\a_2}^{(1)} \nonu \\
& &+ F_1 E_{\a_3}^{(1)}+   F_2 E_{\a_3+\a_4}^{(1)} +
 \bar F_1 E_{-\a_3}^{(1)}+  \bar F_2 E_{-\a_3-\a_4}^{(1)}\nonu \\
(D^{(0)})_{IM}&=& -\pa_x b_1 E_{-\a_1-\a_2}^{(0)} -\pa_x b_2 E_{-\a_2}^{(0)} +
\pa_x \bar b_1 E_{\a_1+\a_2}^{(0)} +\pa_x \bar b_2 E_{\a_2}^{(0)} \nonu\\
& & - \pa_x  F_1 E_{\a_3}^{(0)} -\pa_x F_2 E_{\a_3+\a_4}^{(0)} + \pa_x \bar F_1 E_{-\a_3}^{(0)} +
\pa_x \bar F_2 E_{-\a_3-\a_4}^{(0)}\nonu\\
(D^{(0)})_{Ker}&=&-(b_1 \bar b_1) h_1^{(0)} - (b_1 \bar b_1 +b_2 \bar b_2) h_2^{(0)} + 
(F_1 \bar F_1 +F_2 \bar F_2)h_3^{(0)} + (F_2 \bar F_2)h_4^{(0)} -(b_1\bar b_2) E_{-\a_1}^{(0)} \nonu \\
& &- (b_1 \bar F_1) E_{-\a_1-\a_2-\a_3}^{(0)} -(b_1 \bar F_2) E_{-\a_1-\a_2-\a_3-\a_4}^{(0)} -(b_2 \bar b_1) E_{\a_1}^{(0)}
-(b_2 \bar F_1) E_{-\a_2-\a_3}^{(0)}\nonu \\ 
& &-(b_2 \bar F_2) E_{-\a_2-\a_3-\a_4}^{(0)}
- (\bar b_1 F_1) E_{\a_1+\a_2+\a_3}^{(0)}
-(\bar b_1 F_2) E_{\a_1+\a_2+\a_3+\a_4}^{(0)}- (\bar b_2 F_1) E_{\a_1+\a_2}^{(0)} \nonu \\
& &- (\bar b_2 F_2) E_{\a_2+\a_3 +\a_4}^{(0)}
- (F_1 \bar F_2) E_{-\a_4}^{(0)} - (F_2 \bar F_1) E_{\a_4}^{(0)}\nonu \\
\label{12}
\er
and the equations of motion 
\br
\pa_{t_2}b_1 + \pa^2_{x} b_1 -2 \(b_1\bar b_1+ b_2 \bar b_2 +F_1\bar F_1 +F_2 \bar F_2 \) b_1 =0\nonu \\
\pa_{t_2}b_2 + \pa^2_{x} b_2 -2 \(b_1\bar b_1+ b_2 \bar b_2 +F_1\bar F_1 +F_2 \bar F_2 \) b_2 =0\nonu \\
\pa_{t_2}\bar b_1 - \pa^2_{x} \bar b_1 +2 \(b_1\bar b_1+ b_2 \bar b_2 +F_1\bar F_1 +F_2 \bar F_2 \)\bar  b_1 =0\nonu \\
\pa_{t_2}\bar b_2 - \pa^2_{x} \bar b_2 +2 \(b_1\bar b_1+ b_2 \bar b_2 +F_1\bar F_1 +F_2 \bar F_2 \)\bar  b_2 =0\nonu \\
\pa_{t_2}F_1 + \pa^2_{x} F_1 -2 \(b_1\bar b_1+ b_2 \bar b_2  +F_2 \bar F_2 \) F_1 =0\nonu \\
\pa_{t_2}F_2 + \pa^2_{x} F_2 -2 \(b_1\bar b_1+ b_2 \bar b_2 +F_1\bar F_1  \) F_2 =0\nonu \\
\pa_{t_2}\bar F_1 - \pa^2_{x} \bar F_1 +2 \(b_1\bar b_1+ b_2 \bar b_2  +F_2 \bar F_2 \)\bar  F_1 =0\nonu \\
\pa_{t_2}\bar F_2 - \pa^2_{x} \bar F_2 +2 \(b_1\bar b_1+ b_2 \bar b_2 +F_1\bar F_1  \)\bar  F_2 =0
\label{13}
\er
Consider 
\br
X_0 &=& \eps_{-\a_2-\a_3}E_{\a_2+\a_3}^{(0)}+ \eps_{\a_2+\a_3}E_{-\a_2-\a_3}^{(0)}+
\eps_{-\a_2-\a_3-\a_4}E_{\a_2+\a_3+\a_4}^{(0)}\nonu \\
&+& \eps_{\a_2+\a_3+\a_4}E_{-\a_2-\a_3-\a_4}^{(0)}+
\eps_{-\a_1-\a_2-\a_3}E_{\a_1+\a_2+\a_3}^{(0)}+ \eps_{\a_1+\a_2+\a_3}E_{-\a_1-\a_2-\a_3}^{(0)}\nonu \\
&+&
\eps_{-\a_1-\a_2-\a_3-\a_4}E_{\a_1+\a_2+\a_3+\a_4}^{(0)}+ \eps_{\a_1+\a_2+\a_3+\a_4}E_{-\a_1-\a_2-\a_3-\a_4}^{(0)}
\label{19}
\er
 leading from \rf{18} to the supersymmetry transformations
\br
\d_{X_0}b_1 &=& \eps_{\a_1+\a_2+\a_3+\a_4}F_2 + \eps_{\a_1+\a_2+\a_3} F_1, \nonu \\
\d_{X_0}b_2 &=& \eps_{\a_2+\a_3+\a_4}F_2 + \eps_{\a_2+\a_3} F_1, \nonu \\
\d_{X_0}\bar b_1 &=& \eps_{-\a_1-\a_2-\a_3-\a_4}\bar F_2 + \eps_{-\a_1-\a_2-\a_3} \bar F_1, \nonu \\
\d_{X_0}\bar b_2 &=& \eps_{-\a_2-\a_3-\a_4}\bar F_2 + \eps_{-\a_2-\a_3}\bar  F_1, \nonu \\
\d_{X_0}F_1 &=& -\eps_{-\a_2-\a_3}b_2 - \eps_{-\a_1-\a_2-\a_3} b_1, \nonu \\
\d_{X_0}F_2 &=& -\eps_{-\a_1-\a_2-\a_3-\a_4}b_1 - \eps_{-\a_2-\a_3-\a_4} b_2, \nonu \\
\d_{X_0}\bar F_1 &=& \eps_{\a_2+\a_3} \bar b_2 + \eps_{\a_1+\a_2+\a_3} \bar b_1, \nonu \\
\d_{X_0}\bar F_2 &=& \eps_{\a_1+\a_2+\a_3+\a_4} \bar b_1 + \eps_{\a_2+\a_3+\a_4} \bar b_2, \nonu \\
\label{20}
\er
Of course when $b_1=\bar b_1 =0$ and $F_2 =\bar F_2 =0$ we recover the $p=1$ or $SL(2,1)$ case proposed in 
ref. \cite{aratyn-rasinariu} corresponding to $A_0 = \bar b_2 E_{\a_1} + b_2 E_{-\a_1} + F_1 E_{\a_2}+\bar F_1 E_{-\a_2}$,
 i.e.,
\br
\pa_{t_2}b_2 + \pa^2_{x} b_2 -2 \( b_2 \bar b_2 +F_1\bar F_1  \) b_2 =0\nonu \\
\pa_{t_2}\bar b_2 - \pa^2_{x} \bar b_2 +2 \( b_2 \bar b_2 +F_1\bar F_1  \)\bar  b_2 =0\nonu \\
\pa_{t_2}F_1 + \pa^2_{x} F_1 -2  b_2 \bar b_2    F_1 =0\nonu \\
\pa_{t_2}\bar F_1 - \pa^2_{x} \bar F_1 +2  b_2 \bar b_2  \bar  F_1 =0\nonu \\
\label{13a}
\er
\section{SL(2,1) super-currents}
The components of the sl$(2,1)$ currents 
$J= B^{-1} \pa B$ and $ \bar J= \bar \pa B B^{-1}$ given in equation
\rf{d33} read in terms of the fields defined in \rf{d1} and \rf{d2}:
\br
J_{\a_1}&=& e^{\frac{1}{2}(\varphi_1 +\varphi_2)} \( \pa_z\chi -\frac{1}{2}
\chi(\pa_z\varphi_{1} + \pa_z\varphi_{2}) +\pa_z
f_{1}f_{2} -\frac{1}{2}f_{1}f_{2} \pa_z\varphi_{2} \), \nonu \\
J_{\a_2}&=& e^{-\frac{1}{2}\varphi_1}\(\pa g_2 +\frac{1}{2}g_2\pa \varphi_1 + \pa f_1 \psi -\frac{1}{2} f_1 \pa \varphi_2 \psi - \psi g_2
e^{-\frac{1}{2}(\varphi_1 +\varphi_2)}J_{\a_1}\), \nonu \\
J_{\a_1+\a_2}&=&e^{\frac{1}{2}\varphi_2} \( \pa f_1 -\frac{1}{2} f_1 \pa \varphi_2 - 
g_2 e^{-\frac{1}{2}(\varphi_1 +\varphi_2)}J_{\a_1} \), \nonu \\
J_{\a_2\cdot H}&=& \pa \varphi_2 - \pa f_2 g_2 + \frac{1}{2}f_2 g_2 \pa \varphi_1 - \psi e^{-\frac{1}{2}(\varphi_1 +
\varphi_2)}J_{\a_1}, \nonu \\
J_{(\a_1+\a_2)\cdot H}&=& \pa \varphi_1 + \pa f_1 g_1 - \frac{1}{2}f_1 g_1\pa \varphi_2  - (\psi +g_2g_1) e^{-\frac{1}{2}(\varphi_1 +
\varphi_2)}J_{\a_1}, \nonu \\
J_{-\a_1-\a_2}&=&e^{\frac{1}{2}\varphi_2}\(\pa g_1 -\frac{1}{2}g_1 \pa \varphi_2 -\psi \pa f_2 +\frac{1}{2} \psi \pa \varphi_1 f_2 +
g_1 J_{\a_2\cdot H}\), \nonu \\
J_{-\a_1}&=&e^{-\frac{1}{2}(\varphi_1 +\varphi_2)}\( \pa \psi +\frac{1}{2}\psi (\pa \varphi_1 +\pa \varphi_2)+\psi g_2 \pa f_2 -
\frac{1}{2}\psi  \pa \varphi_1 g_2 f_2 +g_1 e^{\frac{1}{2}\varphi_1}J_{\a_2}  \right. \nonu \\
&-& \left. \psi^2 e^{-\frac{1}{2}(\varphi_1 +\varphi_2)}J_{\a_1}\),
\nonu \\
J_{-\a_2}&=&e^{\frac{1}{2}(\varphi_1}\( \pa f_2 -\frac{1}{2}f_2 \pa \varphi_1 + 
g_1e^{\frac{1}{2}\varphi_1 +\varphi_2)}J_{\a_1}\), \nonu 
\er
and 
\br
\bar J_{-\a_1}&=& e^{\frac{1}{2}(\varphi_1 +\varphi_2)} \( \bar \partial\psi -\frac{1}{2}
\psi(\bar \partial\varphi_{1} + \bar \partial\varphi_{2}) +g_2\bar \partial
g_{1} -\frac{1}{2}g_{2}g_{1} \pa_z\varphi_{2} \), \nonu \\
\bar J_{-\a_2}&=& e^{-\frac{1}{2}\varphi_1}\(\bar \pa f_2 +\frac{1}{2}f_2\bar \pa \varphi_1 + \bar \pa g_1 \chi 
-\frac{1}{2} g_1 \bar \pa \varphi_2 \chi - \chi f_2
e^{-\frac{1}{2}(\varphi_1 +\varphi_2)}\bar J_{-\a_1}\), \nonu \\
\bar J_{-\a_1-\a_2}&=&e^{\frac{1}{2}\varphi_2} \( \bar \pa g_1 -\frac{1}{2} g_1 \bar \pa \varphi_2 - 
f_2e^{-\frac{1}{2}(\varphi_1 +\varphi_2)}\bar J_{-\a_1} \), \nonu \\
\bar J_{\a_2\cdot H}&=& \bar \pa \varphi_2 + \bar \pa g_2 f_2 + \frac{1}{2}f_2 g_2 \bar \pa \varphi_1 - 
\chi e^{-\frac{1}{2}(\varphi_1 + \varphi_2)}\bar J_{-\a_1}, \nonu \\
\bar J_{(\a_1+\a_2)\cdot H}&=& \bar \pa \varphi_1 - \bar \pa g_1 f_1 - \frac{1}{2}f_1 g_1\bar \pa \varphi_2 
 - (\chi +f_1f_2) e^{-\frac{1}{2}(\varphi_1 + \varphi_2)}\bar J_{-\a_1}, \nonu \\
 \bar J_{\a_1+\a_2}&=&e^{-\frac{1}{2}\varphi_2}\(\bar \pa f_1 -\frac{1}{2}f_1 \bar \pa \varphi_2
 -\chi \bar \pa g_2 +\frac{1}{2} \chi \bar \pa \varphi_1 g_2 +
f_1 \bar J_{\a_2\cdot H}\), \nonu \\
\bar J_{\a_1}&=&e^{-\frac{1}{2}(\varphi_1 +\varphi_2)}\( \bar \pa \chi +
\frac{1}{2}\chi (\bar \pa \varphi_1 +\bar \pa \varphi_2)-\chi f_2 \bar \pa g_2 -
\frac{1}{2}\chi  \bar \pa \varphi_1 g_2 f_2 + f_1e^{\frac{1}{2}\varphi_1}\bar J_{-\a_2}  \right. \nonu \\
&-& \left. \chi^2 e^{-\frac{1}{2}(\varphi_1 +\varphi_2)}\bar J_{-\a_1}\),
\nonu \\
\bar J_{\a_2}&=&e^{\frac{1}{2}\varphi_1}\( \bar \pa g_2 -\frac{1}{2}g_2 \bar \pa \varphi_1 +
f_1 e^{-\frac{1}{2}(\varphi_1 +\varphi_2)}\bar J_{-\a_1}\), \nonu \\
\er


\begin{thebibliography}{99}
\bi{kdv-sg}
A. Chodos, Phys. Rev. D {\bf 21}{2818} (1980);
C. A. Tracy and H. Widow, Commun. Math. Phys. {\bf 179} {1} (1996)[solv-int/9506006];
L.A. Ferreira, J. L. Miramontes and J. S. Guill\'{e}n,
Journ. of Math. Phys. {\bf 38} {882} (1997) [hep-th/9606066];
J. Dorfmeister, H. Gradl and J. Szmigielski, {\em Acta Applicandae Math.}
{\bf 53} (1998) 1;
D. Fioravanti and M. Stanishkov, Nucl.\ Phys.\ B {\bf 591}, 685 (2000)
[hep-th/0005158];
\bi{akns-lr}
H.~Aratyn, L.~A.~Ferreira, J.~F.~Gomes and A.~H.~Zimerman,
J.\ Phys.\ A {\bf 33}, L331 (2000) [nlin.si/0007002];
A.M. Kamchatnov and M.V. Pavlov, 
Phys.Lett. A {\bf 301} 269 (2002) [nlin.si/0208025]
\bibitem{Delduc-Gallot}
F.~Delduc and L.~Gallot,
Jour. of Math. Phys. {\bf 39}, 4729 (1998) 
[solv-int/9802013].
\bibitem{Madsen:1999ta}
J.~O.~Madsen and J.~L.~Miramontes,
Commun.\ Math.\ Phys.\  {\bf 217}, 249 (2001)
[hep-th/9905103].
\bibitem{Popowicz:1995jf}
Z.~Popowicz,
J.\ Phys.\ A {\bf 29}, 1281 (1996)
[hep-th/9510185].
\bibitem{Sorin:2002mq}
A.~S.~Sorin and P.~H.~M.~Kersten,
Lett.\ Math.\ Phys.\  {\bf 60}, 135 (2002)
[nlin.si/0201026].
\bibitem{Delduc:2002nd}
F.~Delduc and A.~S.~Sorin,
Proc. of 2002 Workshop on Integrable theories, solitons and duality, eds. L.A. Ferreira, J.F. Gomes and A.H. Zimerman 
, J. High Energy, PRHEP-unesp2002/044, [nlin.si/0206037]
\bibitem{Nissimov:2001cq}
E.~Nissimov and S.~Pacheva,
Jour. of Math. Phys. {\bf 43} 2547 (2002) [nlin.si/0103055]
\bibitem{Ivanov:1999ab}
E.~Ivanov, S.~Krivonos and F.~Toppan,
Mod.\ Phys.\ Lett.\ A {\bf 14}, 2673 (1999)
[solv-int/9912003].
\bibitem{Delduc:1999br}
F.~Delduc, L.~Gallot and A.~Sorin,
Nucl.\ Phys.\ B {\bf 558}, 545 (1999)
[solv-int/9907004].
\bibitem{Delduc:1996mx}
F.~Delduc and L.~Gallot,
Commun.\ Math.\ Phys.\  {\bf 190}, 395 (1997)
[solv-int/9609008].
\bibitem{Ivanov:1996av}
E.~Ivanov and S.~Krivonos,
Phys.\ Lett.\ A {\bf 231}, 75 (1997)
[hep-th/9609191].
\bi{AraRa} H. Aratyn and C. Rasinariu,
Phys.\ Lett.\ B {\bf 391} {99} (1997) [hep-th/9608107];
H. Aratyn, A. Das and C. Rasinariu, Mod. Phys. Lett. A {\bf 12} 2623 (1997)
[solv-int/9704119]
\bi{aratyn-rasinariu} 
H. Aratyn, A. Das, C. Rasinariu and A.H. Zimerman,
in ``{\em Supersymmetry and Integrable Models}'',
H. Aratyn {\sl et.al.} (Eds.), Springer-Verlag, 1998
(Lecture Notes in Physics 502)
\bibitem{Aratyn:2003ym}
H.~Aratyn, J.~F.~Gomes and A.~H.~Zimerman,
Nucl.\ Phys.\ B {\bf 676}, 537 (2004)
[hep-th/0309099];
[hep-th/0409171];
Contribution to the 11-th International Conference on Symmetry Methods in
Physics (SYMPHYS-11), Prague, Czech Republic,
[hep-th/0408231]
\bibitem{ik} T. Inami and H. Kanno, Commun. Math. Phys. {\bf 136} {519} (1991)
\bibitem{kluwer} H. Aratyn, J.F. Gomes, E. Nissimov, S. Pacheva and A.H. Zimerman, ``{\it Symmetry Flows, Conservation Laws and
Dressing Approach to Integrable Models}'' in 
Integrable Hierarchies and Modern
Physical Theories, Ed. H. Aratyn and A. Sorin, Kluwer Academic Publishers (2001), P. 243-275
\end{thebibliography}
\end{document}